\documentclass[a4paper,11pt]{article}
\usepackage{jinstpub} 
\usepackage{subcaption}

\title{\boldmath Drift time calibration of the ultra-low material budget GEM-based TPC for MIXE}

\author[a]{X. Zhao,}
\author[a,*]{M. W. Heiss\note[*]{Corresponding author.},} 
\author[b,*]{F. Garcia,}
\author[a,c]{B. J. Zeh,}
\author[a]{I. Briki,}
\author[d,e]{K. J. Flöthner,}
\author[a]{G. Janka,}
\author[d]{L. Scharenberg,}
\author[c]{B. Banto-Oberhauser,}
\author[d,f]{H. Müller,}
\author[a]{S. Biswas,}
\author[a]{T. Prokscha,}
\author[a]{A. Amato}
\affiliation[a]{PSI Center for Neutron and Muon Sciences CNM, 5232 Villigen PSI, Switzerland}
\affiliation[b]{Helsinki Institute of Physics, University of Helsinki, 00014 Helsinki, Finland}
\affiliation[c]{Institute for Particle Physics and Astrophysics, ETH Zürich, 8093 Zürich, Switzerland}
\affiliation[d]{European Organization for Nuclear Research (CERN), 1211 Geneva, Switzerland}
\affiliation[e]{Helmholtz-Institut für Strahlen-und Kernphysik, University of Bonn, 53115 Bonn, Germany}
\affiliation[f]{Physikalisches Institut, University of Bonn, 53115 Bonn, Germany}

\emailAdd{michael.heiss@psi.ch}
\emailAdd{francisco.garcia@cern.ch}

\abstract{Muon-Induced X-ray Emission (MIXE) is a non-destructive analytical technique that leverages negative muons to probe elemental and isotopic compositions by detecting characteristic muonic X-rays emitted during atomic cascades and gamma rays from nuclear capture processes. By controlling the muon beam momentum, MIXE enables depth-resolved analysis, spanning microns to centimeters, making it ideal for studying compositional variations in fragile, valuable, or operando samples. To enhance its capabilities, we integrated a twin Time Projection Chamber (TPC) tracker with Gas Electron Multiplier (GEM) amplification stages, allowing precise measurement of muon trajectories. A custom-built fiber detector with scintillating fibers and a Silicon Photomultiplier (SiPM) provides permille-level accuracy in drift velocity calibration, essential for accurate spatial reconstruction. This advanced setup correlates muon stopping points with X-ray emissions, paving the way towards element-sensitive imaging and establishing MIXE as a unique tool for high-resolution, depth-specific elemental analysis across diverse scientific applications.}

\keywords{TPC, GEM}

\arxivnumber{2501.10249} 

\begin{document}

\providecommand\hl{}

\maketitle
\flushbottom

\section{Introduction}
\label{sec:intro}

Muon-Induced X-ray Emission (MIXE)~\cite{reidy1978use, kohler1981application, hutson1976tissue, taylor1973observation} is a powerful technique that leverages exotic atom formation by irradiating target samples with a well-defined beam of negative muons. As muons stop within the material, they are captured by target atoms, forming excited atomic states and cascading to the ground state on a picosecond time scale~\cite{Nag2003, Measday2001}. During this process, X-rays are emitted at characteristic energies specific to the atomic species, enabling precise identification of both elemental and isotopic compositions. In addition to atomic cascades, muons can also undergo nuclear capture, a process that becomes more probable in heavier elements~\cite{Suzuki1987}. During nuclear capture, the muon induces the conversion of a proton into a neutron, resulting in a transmuted nucleus. These nuclei de-excite by emitting a combination of nucleons and gamma rays. The combination of muonic X-rays from atomic cascades and gamma rays from nuclear interactions provides multiple pathways for elemental identification. High-purity germanium (HPGe) detectors are used to measure these emissions with excellent energy resolution and efficiency over a broad spectrum, ensuring accurate and detailed analysis. One significant advantage of MIXE over other non-destructive techniques, like X-ray fluorescence (XRF)~\cite{Philip1992}, is its depth sensitivity. By adjusting the momentum of implanted muons, their stopping range is well defined, spanning micrometers to centimeters, enabling advanced non-destructive, depth-resolved elemental analysis. Thus, MIXE is highly suitable for assessing valuable, rare, or fragile samples, making it ideal for studying cultural heritage artifacts~\cite{Beda2023} or for operando investigations such as battery charging kinetics~\cite{Edouard2025}. To further enhance the method, we aim to correlate the precise muon stopping location with the corresponding X-ray emissions, thereby enabling a novel, element-sensitive tomographic technique. 
\begin{figure}[htbp]
  \begin{subfigure}{.35\textwidth}
  \centering
  \includegraphics[width=0.9\textwidth,clip]{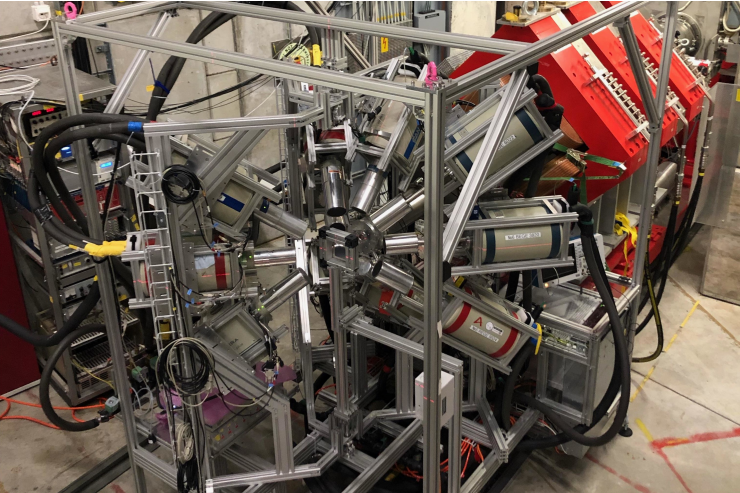}
  \label{fig:sgiant}
\end{subfigure}%
  \begin{subfigure}{.3\textwidth}
  \centering
  \includegraphics[width=0.77\textwidth,clip]{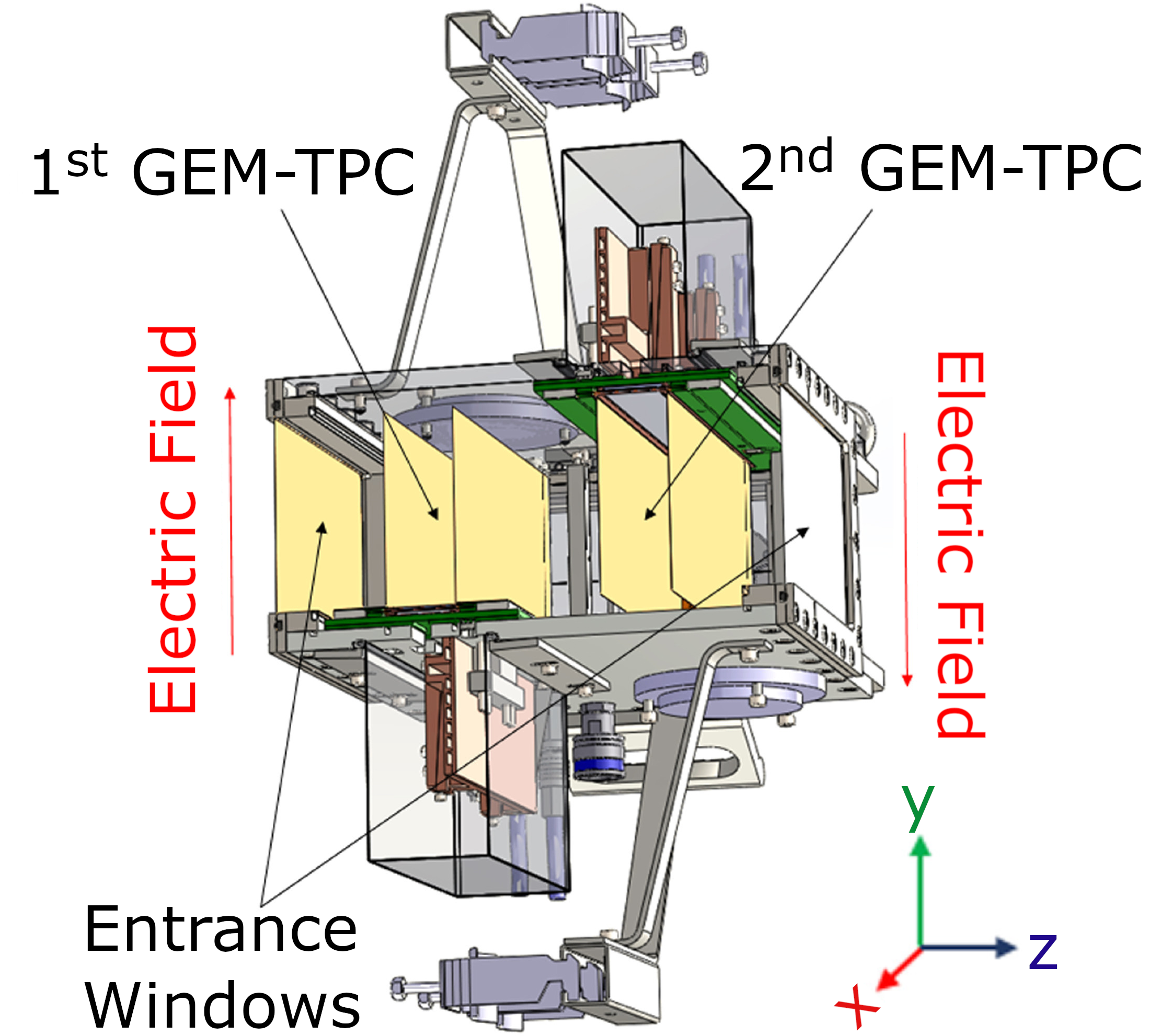}
  \label{fig:stracker1}
  \end{subfigure}%
  \begin{subfigure}{.35\textwidth}
  \centering
  \includegraphics[width=0.8\textwidth,clip]{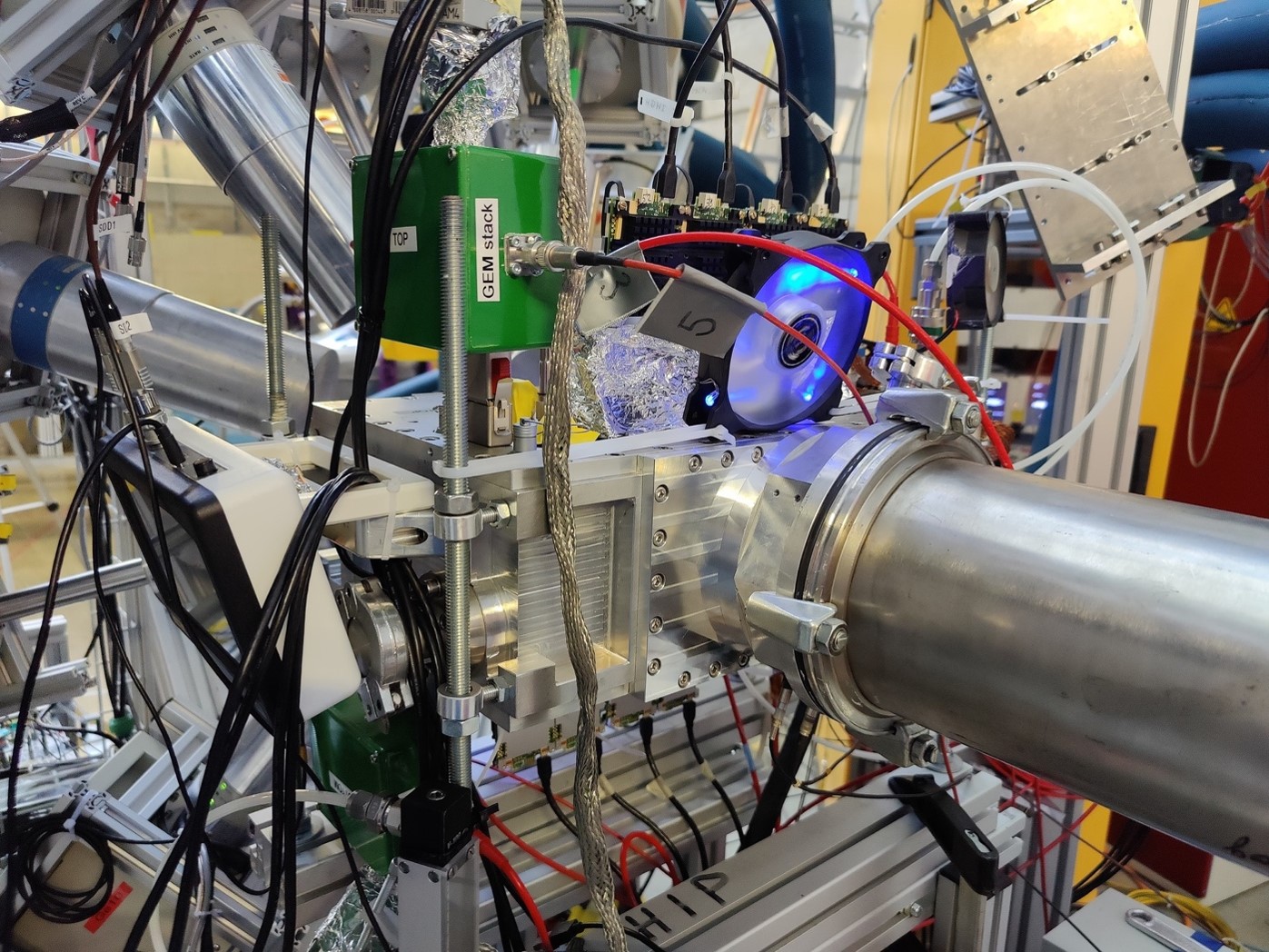}
  \label{fig:stracker2}
  \end{subfigure}%
\caption{\hl{The GIANT instrument for MIXE, installed at the $\pi$E1 beamline area at PSI (left). The schematic of the twin GEM-TPC with both GEM-TPCs housed in a single vessel and operating with opposite electric field directions (middle). The tracker mounted to the beamport of $\pi$E1, upstream of the target to track incoming muons (right). }\label{fig:mixe_setup}}
\end{figure}

\hl{Supporting this development, we adapted a twin Time Projection Chamber (TPC) with Gas Electron Multiplier (GEM)~{\cite{GARCIA201818}}. In the MIXE experimental setup at the Paul Scherrer Institute (PSI), the elemental fingerprint spectra from the emitted characteristic X-rays and gamma rays are acquired using GIANT (Germanium Array for Non-destructive Testing)~{\cite{Lars2023}}, which is positioned around the target (Figure~{\ref{fig:mixe_setup}}, left). As shown in the middle of Figure~{\ref{fig:mixe_setup}}, the tracker consists of two GEM-TPCs housed within a single vessel, with the first TPC oriented to drift upward and the second TPC downward. The X-position of the muon is determined by the projection of charge clusters onto the strips, whereas the Y-position is derived from the drift time of the electrons~{\cite{Garcia_2025}}. }

\begin{figure}[htbp]
\begin{subfigure}{.49\textwidth}
  \centering
  \includegraphics[width=0.9\textwidth,clip]{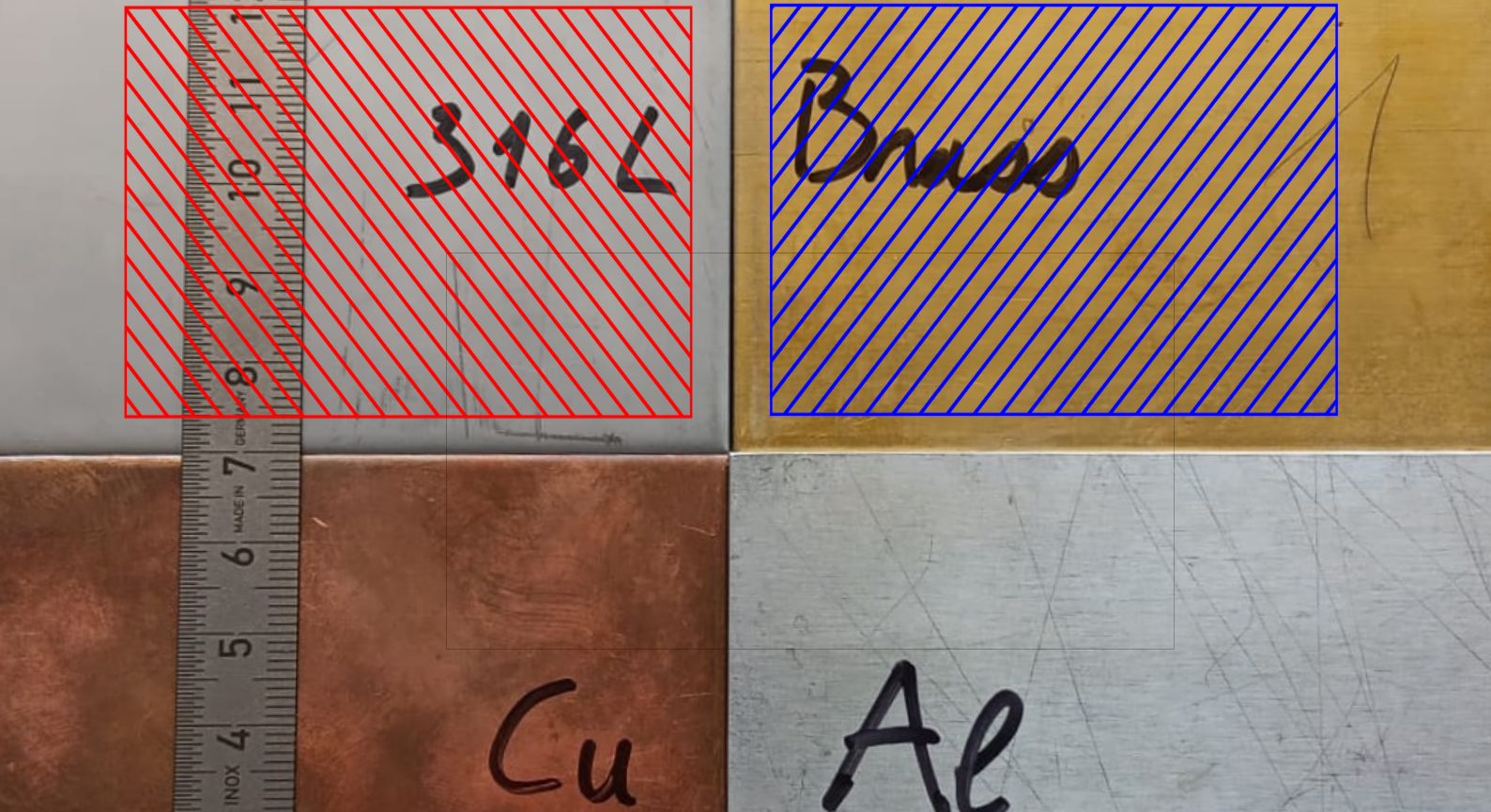}
  \label{fig:1}
\end{subfigure}%
\begin{subfigure}{.49\textwidth}
  \centering
  \includegraphics[width=0.95\textwidth,clip]{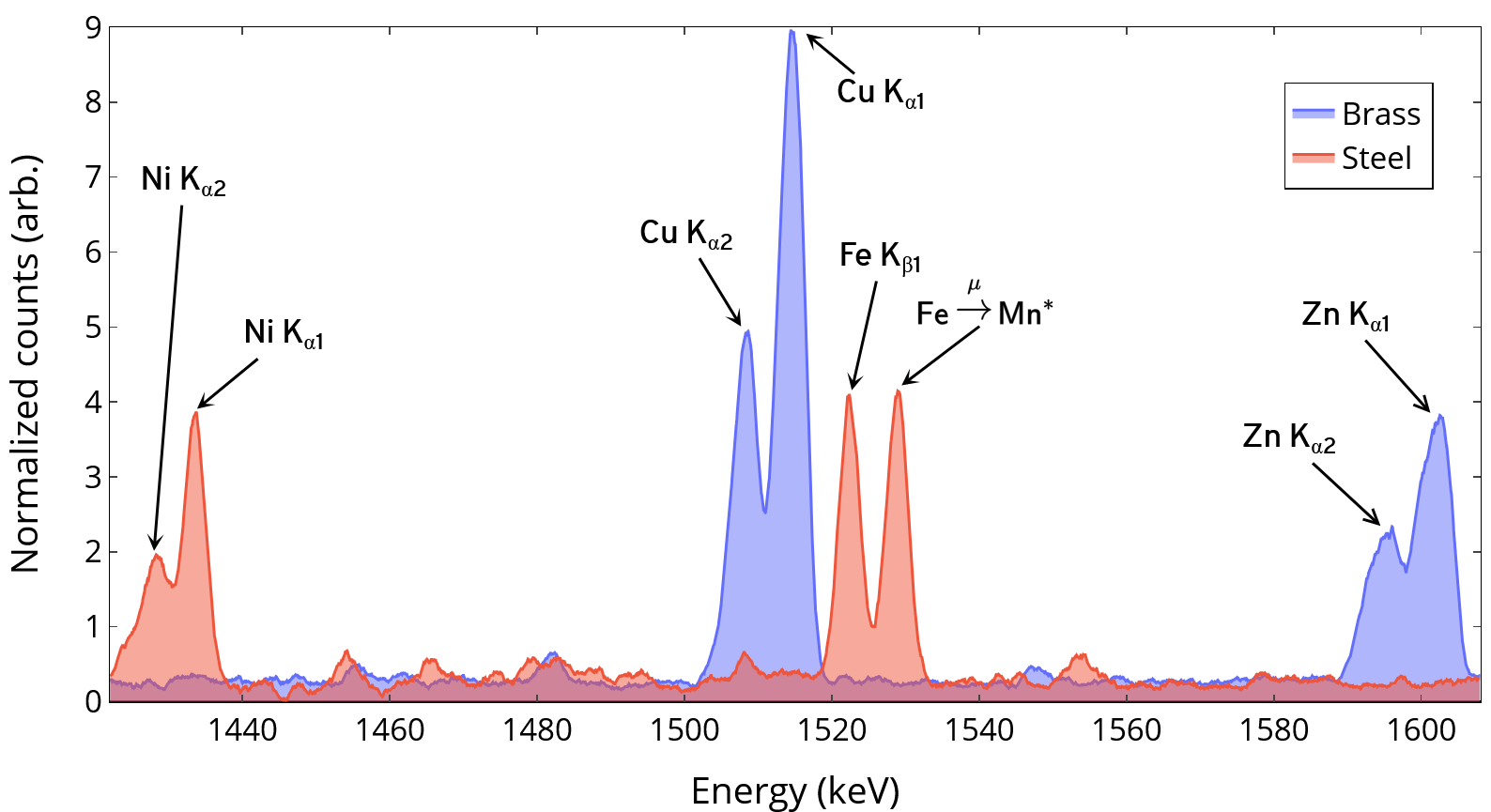}
  \vspace{-0.2cm}
  \label{fig:2}
\end{subfigure}%
\caption{\hl{Photo of the multi-material target, composed of stainless steel, brass, copper and aluminum plates (left). Muonic X-ray spectrum obtained by selecting regions within the brass and stainless steel plates (right). The $K_\alpha$ lines from Cu and Zn in brass, as well as the $K_\alpha$ line from Ni and the $K_\beta$ line from Fe in stainless steel are clearly distinguishable through spatial selection on the target.}\label{fig:spectrum}}
\end{figure}

\hl{A proof-of-concept demonstration of elemental imaging was carried out using 60 MeV/$c$ muons and the tracker operated with Ar/CO$_2$ (75:25) at the $\pi$E1 beamline in June 2023. The tracker was used to record precise muon trajectories, from which their stopping positions within the target were reconstructed. As shown in the left of Figure~{\ref{fig:spectrum}}, the target consisted of four material plates. By combining the stopping X/Y position information provided by the tracker with the characteristic X-ray spectra recorded by the HPGe detectors, clear separation of different materials in the sample was achieved. }

\hl{Since the TPC was filled with a Ar/CO$_2$ (75:25) gas mixture, the spatial resolution of MIXE was primarily limited by multiple scattering within the tracker. To mitigate this effect, an ultra-low-mass He/CO$_2$ (90:10) mixture was employed during both the DRD1 test beam at the H4 beamline of the SPS at CERN in 2024~{\cite{Garcia_2025}}, and the full MIXE campaign in June 2024 (analysis ongoing). Reliable position reconstruction requires accurate calibration of the drift velocity in the gas. As no external reference system was available at PSI and in situ calibration was required, a compact fiber detector was developed and successfully used to perform the calibration, achieving permille-level precision. Given the intrinsic twin GEM-TPC resolution of $\sim 200\,\mathrm{\mu m}$~{\cite{Garcia_2025}}, calibration precision on the order of $4\,\text{\textperthousand}$ is required to not limit the absolute position precision at the edges of the active volume. The design of this detector is presented in Section~{\ref{sec:chap_2}}, while results of the drift time calibration for both gas mixtures are presented in Section~{\ref{sec:chap_3}}.}

\section{Drift time calibration}\label{sec:chap_2}

\begin{figure}[htbp]
\begin{subfigure}{.4\textwidth}
  \centering
  \includegraphics[width=0.9\textwidth,clip]{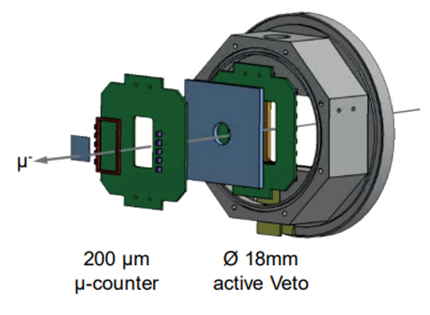}
  \label{fig:sfig1}
\end{subfigure}%
\begin{subfigure}{.3\textwidth}
  \centering
  \includegraphics[width=0.7\textwidth,clip]{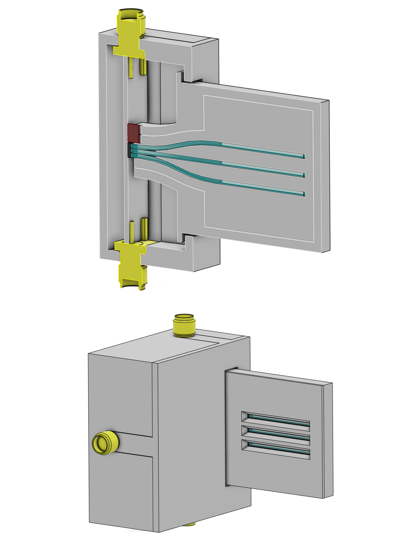}
  \label{fig:sfig2}
  \end{subfigure}%
\begin{subfigure}{.3\textwidth}
  \centering
  \includegraphics[width=0.8\textwidth,clip,trim={20 20 30 90}]{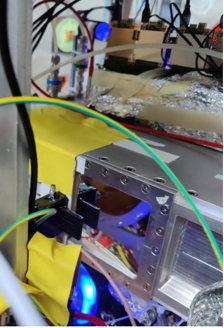}
  \label{fig:sfig3}
\end{subfigure}%
\caption{\hl{Schematic of the entrance detector, comprising a $\mu$-counter paired with a veto detector (left).} Construction drawings of the fiber detector, showing a full frontal view (middle top) and a cutaway view (middle bottom). The fiber detector mounted in front of the upstream window of the TPC during operation (right).\label{fig:optical_fiber_detector}}
\end{figure}

The optical fiber detector was placed in front of the exit window of the twin GEM-TPC to simultaneously detect incoming muons in multiple systems: the entrance detector~\cite{Lars2023}, one of the fibers in the optical fiber detector, and the tracker. 
\hl{Incoming muons are first registered by the entrance detector. As shown in Figure~{\ref{fig:optical_fiber_detector}}, it composed of a $20 \times 20\,\mathrm{mm^2}\times 200\,\mathrm{\mu m}$ BC-400 plastic scintillator, masked with a $10 \times 10\,\mathrm{cm^2}\times 8\,\mathrm{mm}$ BC-400 veto counter with a $18\,\mathrm{mm}$ central hole. The scintillators are paired with a series of SiPMs (ASD-NUV3S-P type by AdvanSiD) and in-house developed pre-amplifiers~{\cite{Stella_thesis}}, providing a nanosecond-level time resolution.} The fiber detector (Figure~\ref{fig:optical_fiber_detector}, center) consists of a precision 3D printed mounting structure ($35\,\mathrm{\mu m}$ resolution), housing three scintillating optical fibers of approximately $50\,\mathrm{mm}$ length, with a $1 \times 1 \,\mathrm{mm}^2$ cross section, spaced exactly $4\,\mathrm{mm}$ apart. The active areas of the fibers are shielded from light by two thin ($10\,\mathrm{\mu m}$) titanium foils, allowing low energy muons to enter with minimal energy loss. The rest of the housing is sufficiently thick that muons are absorbed in the material before reaching the fibers. The signal is read out by a fast SiPM mounted on an optimized readout board (onsemi MICROFJ-SMA-60035-GEVB), utilizing the fast signal output, which is amplified by two Minicircuits ZFL-1000LN+ amplifiers in series. 

With the entrance detector, the timestamp of incoming muons was recorded and considered as the $t_0$ of each event. Thus, a stringent cut on the total drift time sum in both GEM-TPCs allows selecting parallel tracks. During data analysis, events in the GEM-TPCs that correspond to hits in one of the three fibers are identified by coincidence. This correlation enables the generation of a drift time distribution with three distinct peaks, each representing the interaction of a muon with one of the fibers.
The peaks are then fitted with a sum of three coupled Gaussian functions:
\begin{equation}
    f(x) = A_1\exp\left(-\frac{(x-C-D)^{2}}{W_1}\right)+A_2\exp\left(-\frac{(x-C)^{2}}{W_2}\right)+A_3\exp\left(-\frac{(x-C+D)^{2}}{W_3}\right),
    \label{eq:fit}
\end{equation}
where the model assumes that the Gaussian peaks are symmetrically distributed around the central time $C$, with offset timing $D$. The parameters $A_1, ~A_2$ and $A_3$ represent the amplitudes and $W_1$, $W_2$ and $W_3$ represent the widths of the respective distributions. Since the separation between the fibers directly corresponds to the offset timing, it is then simply given by $v_{d} = 4\,\mathrm{mm} / {\left| D \right|}$.

\section{Results}\label{sec:chap_3}

During the PSI beam experiments, the tracker was evaluated using two gas mixture: Ar/CO$_2$ (75:25), and He/CO$_2$ (90:10). High voltages were applied to the GEM-TPCs, with 300 V/cm for the drift regions and 3.01 kV (top) and 3 kV (bottom) applied to the triple GEM stacks for Ar/CO$_2$, and 325 V/cm for the drift regions and 2.16 kV (top) and 2.15 kV (bottom) applied to the triple GEM stacks for He/CO$_2$.

\begin{figure}[htbp]
\begin{subfigure}{.5\textwidth}
  \centering
  \includegraphics[width=0.95\textwidth,clip]{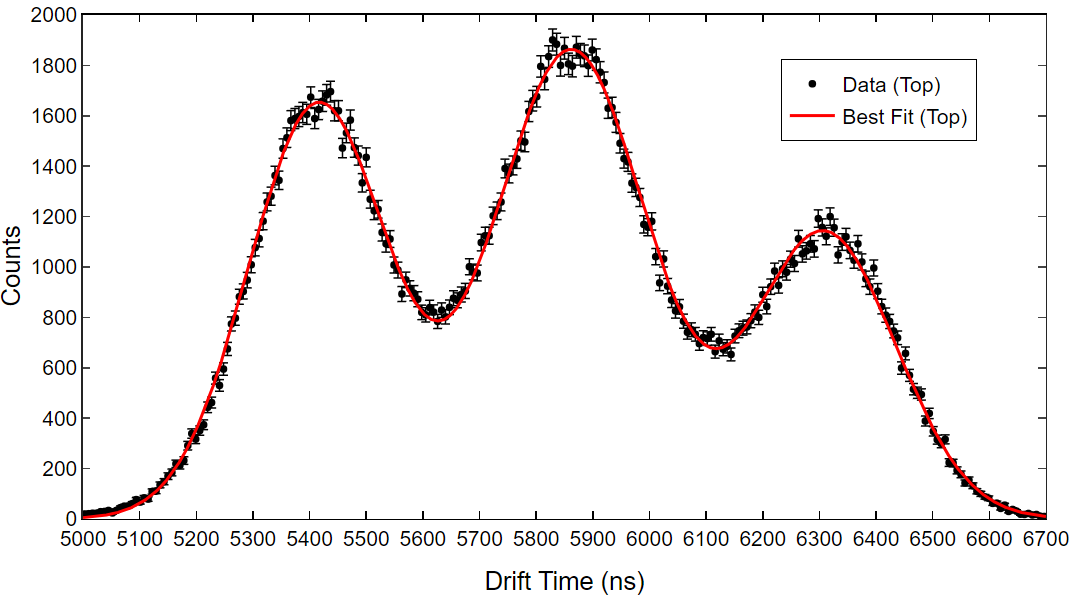}
\end{subfigure}%
\begin{subfigure}{.5\textwidth}
  \centering
  \includegraphics[width=0.95\textwidth,clip]{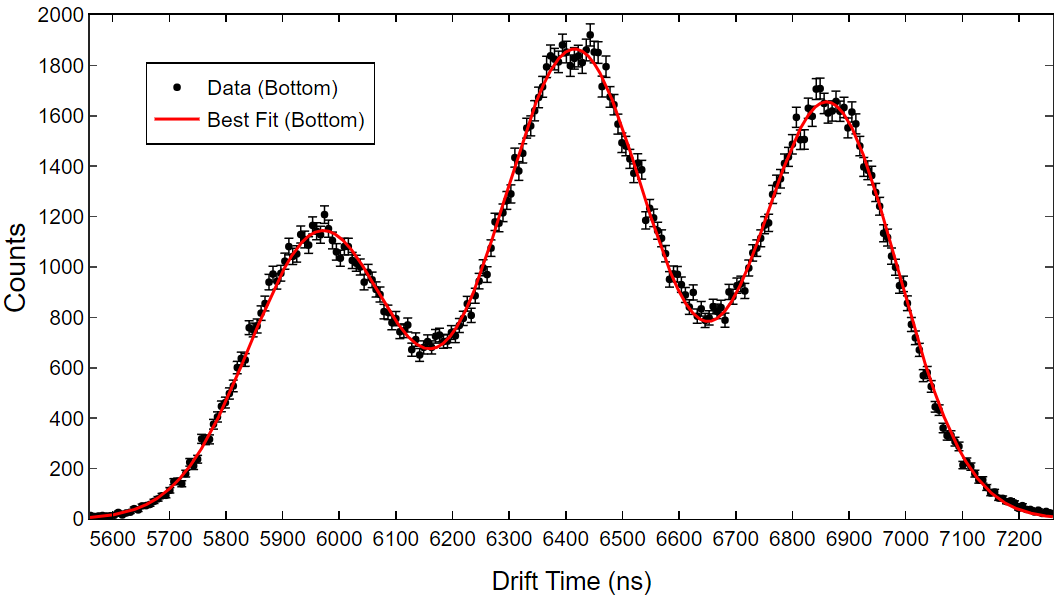}
  \end{subfigure}%
\caption{The drift time distribution and best fit of the Top GEM-TPC (left) and the Bottom GEM-TPC (right) using Ar/CO2 (75:25) and the fiber detector during the September 2023 beamtime.\label{fig:result_ar}}
\end{figure}

\begin{figure}[htbp]
\begin{subfigure}{.5\textwidth}
  \centering
  \includegraphics[width=0.95\textwidth,clip]{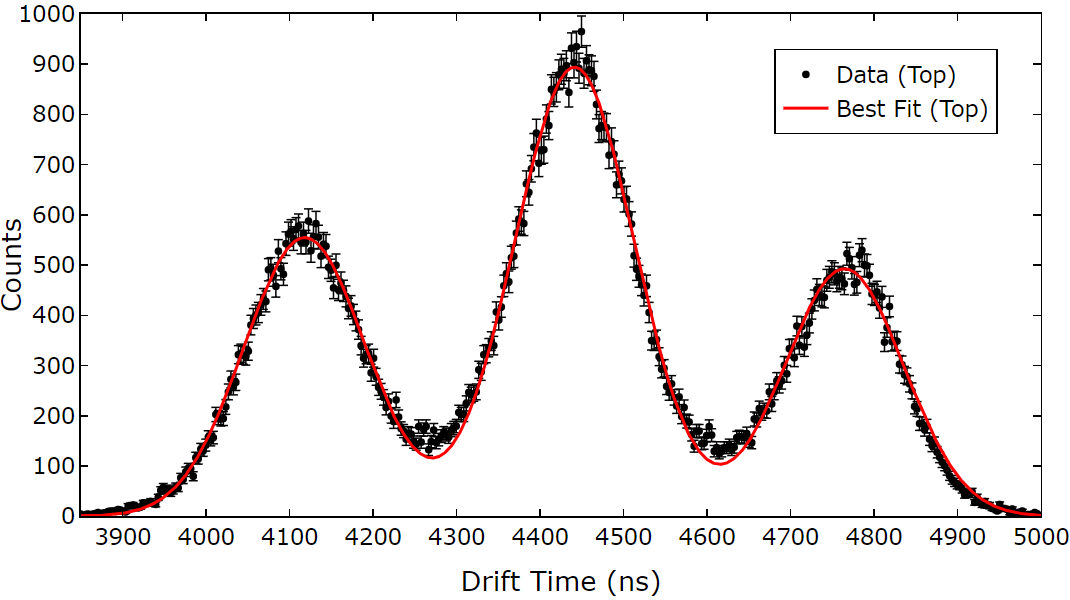}
\end{subfigure}%
\begin{subfigure}{.5\textwidth}
  \centering
  \includegraphics[width=0.95\textwidth,clip]{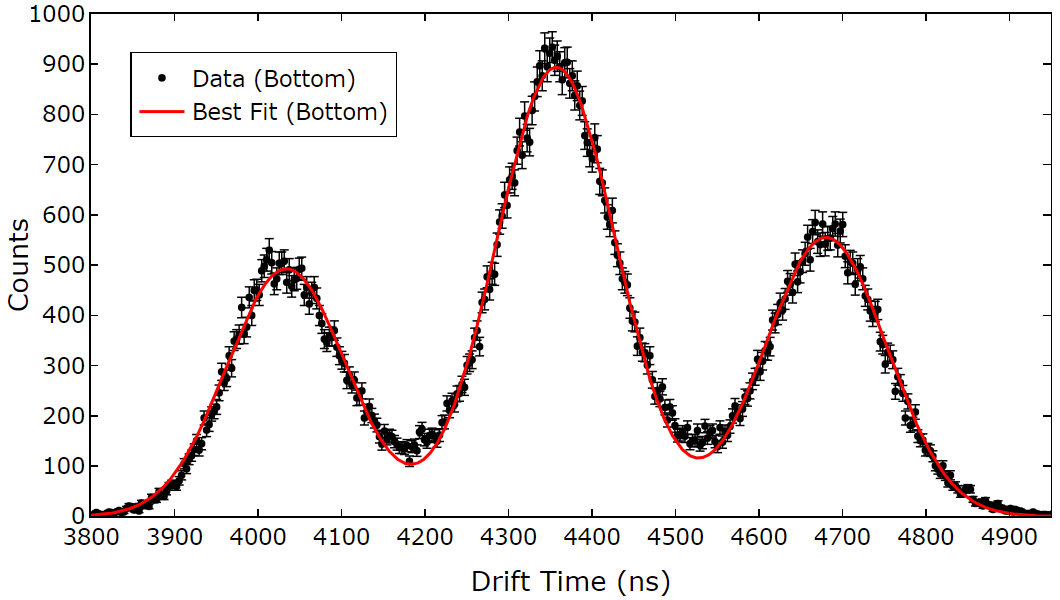}
  \end{subfigure}%
\caption{The drift time distribution and best fit of the Top GEM-TPC (left) and the Bottom GEM-TPC (right) using He/CO2 (90:10) and the fiber detector during the June 2024 beamtime.\label{fig:result_he}}
\end{figure}

Utilizing a muon beam with momenta of 50 (He/CO$_2$) to 60 (Ar/CO$_2$) MeV/c, the drift times for both the top and bottom GEM-TPCs were evaluated, as depicted in Figure~\ref{fig:result_ar} and Figure~\ref{fig:result_he}. Using equations described in Section~\ref{sec:chap_2}, the drift velocities for the Ar/CO$_2$ gas mixture were calculated to be $v^\text{top}_{d} = (8.956 \pm 0.013)\,\mathrm{mm/\mu s}$ and $v^\text{bottom}_{d} = (8.954 \pm 0.012)\,\mathrm{mm/\mu s}$, respectively. For the He/CO$_2$ gas mixture, the drift velocities were similarly calculated to be $v^\text{top}_{d} = (12.376 \pm 0.011)\,\mathrm{mm/\mu s}$ and $v^\text{bottom}_{d} = (12.375 \pm 0.011)\,\mathrm{mm/\mu s}$.

This in-situ calibration method provides an efficient and precise approach to determine the tracker’s drift velocity, eliminating the need for additional large-scale detectors. The measurements are consistent with results obtained with high energy muons using a reference tracking system~\cite{Garcia_2025}. This capability will be crucial to ensure reliable operation while being able to change the parameters of the GEM-TPCs to cover the range of muon momenta required for depth-selective studies.

\acknowledgments
This work is based on experiments performed in September 2023 and June 2024 at the PiE1 beamline of the Swiss Muon Source S$\mu$S (proposals 20222993 and 20231296), Paul Scherrer Institute, Villigen, Switzerland.
The authors gratefully acknowledge the Laboratory for Particle Physics (LTP) and the Laboratory for Muon Spin Spectroscopy (LMU) at PSI for their generous support. We also extend our gratitude to members of the CERN GDD team and of RD51/DRD1 test beam working group for their invaluable contributions.
Many thanks to C. Kaya, J. Kunkel, B. Voss, H. Risch, C. Schmidt, R. Turpeinen, J. Heino, S. Rinta-Antila, R. de Oliveira, and the CERN Micro Pattern Technology (MPT) Workshop for their support during the design and assembly of the TPC. This work was supported by the SNSF bilateral program grant 214052, SNSF Sinergia grant 193691 and the SDSC collaborative project SAMURAI (C22-12L).


\begin{thebibliography}{99}

\bibitem{reidy1978use}
Reidy et al.,
\emph{Use of muonic x-rays for nondestructive analysis of bulk samples for low Z constituents},
Anal. Chem. {\bf 50} (1978) 40-44.

\bibitem{kohler1981application}
K{\"o}hler et al.,
\emph{Application of muonic X-ray techniques to the elemental analysis of archeological objects},
Nucl. Instrum. Methods Phys. Res. {\bf 187} (1981) 563-568.

\bibitem{hutson1976tissue}
Hutson et al.,
Tissue chemical analysis with muonic x rays,
Radiology (1976).

\bibitem{taylor1973observation}
Taylor et al.,
\emph{Observation of Muonic X-rays from Bone},
Radiat. Res. (1973).

\bibitem{Nag2003}
Nagamine, K.,
\emph{Introductory muon science},
Cambridge University Press (2003).

\bibitem{Measday2001}
D.F. Measday,
\emph{The nuclear physics of muon capture},
Phys. Rep. {\bf 354} (2001) 243-409.

\bibitem{Suzuki1987}
Suzuki T. et al. ,
\emph{Total nuclear capture rates for negative muons},
Phys. Rev. C {\bf 35} (1987) 2212--2224.

\bibitem{Philip1992}
Philip J. Potts et al. ,
\emph{X-ray fluorescence spectrometry},
J. Geochem. Explor. {\bf 44} (1992) 251-296.

\bibitem{Beda2023}
Beda A. Hofmann et al. ,
\emph{An arrowhead made of meteoritic iron from the late Bronze Age settlement of Mörigen, Switzerland and its possible source},
J. Archaeol. Sci. {\bf 157} (2023).

\bibitem{Edouard2025}
Q. Edouard et al.,
\emph{Overcoming the probing-depth dilemma in spectroscopic analyses of batteries with muon-induced X-ray emission (MIXE)},
J. Mater. Chem. A (2025).

\bibitem{GARCIA201818}
F. García et al.,
\emph{A GEM-TPC in twin configuration for the Super-FRS tracking of heavy ions at FAIR},
NIM-A {\bf 884} (2018) 18-24.

\bibitem{Lars2023}
Lars Gerchow et al.,
\emph{GermanIum array for non-destructive testing (GIANT) setup for muon-induced x-ray emission (MIXE) at the Paul Scherrer Institute},
Rev. Sci. Instrum. (2023).

\bibitem{Garcia_2025}
F. García et al.,
\emph{The ultra-low material budget GEM-based TPC for tracking with VMM3a readout},
JINST (2025) C04004.

\bibitem{Stella_thesis}
Vogiatzi, Stergiani Marina,
\emph{Studies of muonic 185,187Re, 226Ra, and 248Cm for the extraction of nuclear charge radii},
Doctoral Thesis. (2023).

\end{thebibliography}
\end{document}